\documentclass[sigconf]{acmart}

\setcopyright{acmcopyright}
\usepackage{graphicx}
\usepackage{lscape}
\usepackage{upgreek}
\usepackage{caption} 
\usepackage{hyperref}
\usepackage{todonotes}
\usepackage{lineno}
\usepackage{url}
\usepackage{array}
\usepackage{enumitem}
\usepackage{multirow}
\usepackage{soul}
\usepackage{float}
\usepackage{color}
\usepackage{pgfplots}
\usepackage{booktabs}
\usepackage[export]{adjustbox}
\usepackage{colortbl}
\usepackage{fontawesome}
\usepackage[normalem]{ulem}
\acmDOI{xx.xxx/xxx_x}

\acmISBN{xxx-x-xxxx-xxxx-x/YY/MM}

\acmConference[XXX 2024]{XXX Conference}{Month 2024}{City, Country}
\acmYear{2024}
\copyrightyear{2024}

\acmArticle{4}
\acmPrice{15.00}

\begin{document}
\title{6G Software Engineering: A Systematic Mapping Study}
%\titlenote{Produces the permission block, and copyright information}
% \subtitle{Extended Abstract}
%\subtitlenote{The full version of the author's guide is available as \texttt{acmart.pdf} document}
  
 %\renewcommand{\shorttitle}{SIG Proceedings Paper in LaTeX Format}

\author{Ruoyu Su}
% % \authornote{Dr.~Trovato insisted his name be first.}
% %\orcid{1234-5678-9012}
\affiliation{
\institution{M3S, University of Oulu}
%   %\streetaddress{P.O. Box 1212}
   \city{Oulu} 
%   %\state{Ohio} 
   \country{Finland}
%   %\postcode{43017-6221}  
 }
 \email{ruoyu.su@.oulu.fi}

 \author{Xiaozhou Li}
% % \authornote{The secretary disavows any knowledge of this author's actions.}
 \affiliation{
   \institution{M3S, University of Oulu}
%   %\streetaddress{P.O. Box 1212}
  \city{Oulu} 
%   %\state{Ohio} 
   \country{Finland}
%   %\postcode{43017-6221}  
 }
 \email{xiaozhou.li@oulu.fi}

  \author{Davide Taibi}
% % \authornote{The secretary disavows any knowledge of this author's actions.}
 \affiliation{
   \institution{M3S, University of Oulu}
%   %\streetaddress{P.O. Box 1212}
  \city{Oulu} 
%   %\state{Ohio} 
   \country{Finland}
%   %\postcode{43017-6221}  
 }
 \email{davide.taibi@oulu.fi}

 %\author{Davide Taibi}
% % \authornote{This author is the one who did all the really hard work.}
 %\affiliation{
   %\institution{M3S, University of Oulu}
%   %\streetaddress{1 Th{\o}rv{\"a}ld Circle}
   %\city{Oulu} 
   %\country{Finland}~\\
 %\institution{CloudSEA.AI, Tampere University}
  %\city{Tampere}
 % \country{Finland}
 % }
 %\email{davide.taibi@oulu.fi}

%\affiliation{%
 % \institution{Inria Paris-Rocquencourt}
 % \city{Rocquencourt}
 % \country{France}
%\affiliation{%
% \institution{Rajiv Gandhi University}
 %\state{Arunachal Pradesh}
 %\country{India}}
%\affiliation{%
 % \institution{Tsinghua University}
  %\streetaddress{30 Shuangqing Rd}
  %\city{Haidian Qu} 
  %\state{Beijing Shi}
  %\country{China}
%}

%\affiliation{%
 % \institution{Palmer Research Laboratories}
 % \streetaddress{8600 Datapoint Drive}
  %\city{San Antonio}
  %\state{Texas} 
  %\country{USA}
 % \postcode{78229}}  

\renewcommand{\shortauthors}{R. Su et al.}

\begin{abstract}
%This paper provides a sample of a \LaTeX\ document which conforms,
%somewhat loosely, to the formatting guidelines for
%ACM SIG Proceedings.%\footnote{This is an abstract footnote}
6G will revolutionize the software world allowing faster cellular communications and a massive number of connected devices. 6G will enable a shift towards a continuous edge-to-cloud architecture. Current cloud solutions, where all the data is transferred and computed in the cloud, are not sustainable in such a large network of devices. Current technologies, including development methods, software architectures, and orchestration and offloading systems, still need to be prepared to cope with such requirements. In this paper, we conduct a Systematic Mapping Study to investigate the current research status of 6G Software Engineering. Results show that 18 research papers have been proposed in software process, software architecture, orchestration and offloading methods. Of these, software architecture and software-defined networks are respectively areas and topics that have received the most attention in 6G Software Engineering. In addition, the main types of results of these papers are methods, architectures, platforms, frameworks and algorithms. For the five tools/frameworks proposed, they are new and not currently studied by other researchers. The authors of these findings are mainly from China, India and Saudi Arabia. The results will enable researchers and practitioners to further research and extend for 6G Software Engineering.
\end{abstract}

\keywords{6G, software engineering, systematic mapping study, edge computing}

\maketitle

\section{Introduction}

With the rapid development of communication applications, communication technology is experiencing revolutionary evolution from generation to generation~\cite{wang2023road}. A generation of mobile communication systems is updated almost every decade.~\cite{chen2020vision}. Up to now, cellular mobile communication systems have developed through five generations (1G to 5G). Starting in 2019, 5G has been officially commercialized and is now widely used worldwide~\cite{dang2020should}. Although further enhancements to 5G are still underway, academics and industry have begun research into the next generation, namely 6G.

6G is the Sixth Generation standard for cellular communications and is presently in development to succeed 5G~\cite{nguyen2021security}. 6G will support faster speeds and expand connectivity in areas traditionally covered by 5G applications, enabling the creation of extremely large edge-to-cloud continuous systems~\cite{chen2020vision}. In addition, 6G will create new software-specific challenges both for the development of the 6G products and applications and services utilizing 6G networks. The number of connected devices will explode and the current architectures, orchestration, and scalability methods and tools are not capable of supporting such complex, heterogeneous, and highly distributed 6G software systems. In addition, the application of current technologies will create a very high energy overhead, due to the lack of optimization for such a large number of connected devices~\cite{hakeem2022vision}.

Current orchestration systems, such as Kubernetes\footnote{Kubernetes: https://kubernetes.io} and OpenShift
\footnote{RedHat Open Shift: https://www.redhat.com/en/technologies/cloud-computing/openshift}
are designed to scale the computation horizontally, distributing the computation to different nodes. As an example, Kubernetes can scale up to 5,000 computation nodes while using a very large amount of energy for scaling the tasks on top of the energy used by each node~\cite{hakeem2022vision}. Cloud-based orchestration tools assume that the “nodes” (i.e., containers) are created for the application’s needs. In 6G infrastructure, the nodes are often tied to physical devices; thus, the application code needs to adapt the infrastructure instead of creating it. Based on the location of the computation and the need for data transfer, orchestration models require an optimization layer considering not only the data transfer but also the availability of green energy, enabling the execution of tasks as closely as possible to the data source, or using as much green energy as possible. For this reason, there is a need to develop new energy-efficient scalability and orchestration models and tools. Software architectures should be also designed to allow fast scaling.

To overcome the aforementioned issues, %Business Finland funded 6GSoft, a 3-years research project
we aimed at extend or developing software development methods, energy-Aware 6G orchestration models, architectural support and to identify business-driven 6G software development models through a nationally funded project. 

In this paper we report the first step of this project, identifying the current research status in Software Engineering practices applied in the context of 6G. In particular, we are interested in papers on software process, common software architecture, orchestration and offloading methods applied to 6G. Based on our goal, we are working to investigate the main topics, studies, results and authors in 6G Software Engineering, especially focusing on the areas and topics that have received the most attention and tools/frameworks that are most studied. For this purpose, we conducted a Systematic Mapping Study of the scientific literature~\cite{petersen2015guidelines}, identifying 18 papers in total out of 479. 

In the remainder of this paper we refer to software process, common software architecture, orchestration and offloading methods applied to 6G as "\textbf{6G Software Engineering}".

% The contributions of the paper are that the results of our work 
The results of this work can inform researchers about the existing research status in this domain so that they can have a basis when continuing research in this field.

% At the same time, an improved understanding of what kinds of results are obtained, especially in software process, software architecture, orchestration and offloading methods and business-driven software development, may be useful to practitioners interested in 6G Software Engineering.

The remainder of this paper is organized as follows. Section 2 reports on the research method employed to conduct the systematic mapping study. In Section 3, we analyze the results that address the goal of the study. Section 4 presents main discussion points according to our analysis and Section 5 researches the threats to validity. Finally, Section 6 summarizes the paper the outlines our future works.

\section{Methodology}

The goal of the systematic mapping study is to investigate the current research status of 6G Software Engineering within the confines of the funded project. In the context of our research, we identified three main research questions:
% that aim to target the problems under the perspectives of the topics/studies and obtained the results involved.

% These are listed in the following:

\textbf{RQ$_1$ Main topics and studies in 6G Software Engineering}
\begin{itemize}

    \item \textbf{RQ$_{1.1}$}. \emph{How many studies have been proposed in the context of 6G Software Engineering? (Software Process, Software Architecture, Orchestration, Offloading)}

    \item \textbf{RQ$_{1.2}$}. \emph{Which areas and topics have received the most attention in 6G Software Engineering?}
    
\end{itemize}

\textbf{RQ$_2$ Main results obtained in 6G Software Engineering}

\begin{itemize}

    \item \textbf{RQ$_{2.1}$}. \emph{What are the main results reported?}

    \item \textbf{RQ$_{2.2}$}. \emph{Which tools/frameworks are most being studied?}
    
\end{itemize}

\textbf{RQ$_3$ Main authors and distribution in 6G Software Engineering}

\begin{itemize}

    \item \textbf{RQ$_{3.1}$}. \emph{Who are the main researchers in 6G Software Engineering?}

    \item \textbf{RQ$_{3.2}$}. \emph{Given the main knowledge areas, how are these researchers distributed?}

\end{itemize}

Our systematic mapping study follows the commonly adopted guidelines provided by Peterson et al.~\cite{petersen2015guidelines}. Furthermore, we followed the guidelines of Wohlin~\cite{wohlin2014guidelines}, which uses the approach called "snowballing", i.e., analyzing references of major studies in order to extract more related information when summarizing the existing knowledge on a specific subject.

\subsection{Search Strategy}

\begin{figure}[h]
    \centering
    \includegraphics[width=\linewidth]{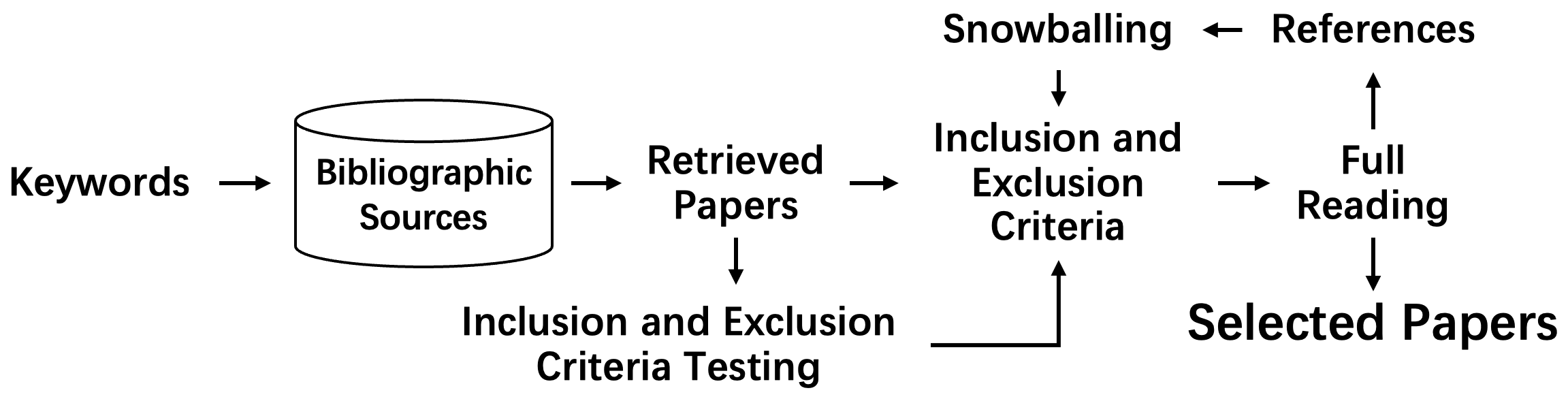}
    \caption{The Search and Selection Process}
    \label{fig:sasprocess}
\end{figure}

The search strategy is shown in Fig.~\ref{fig:sasprocess}. We first defined the search terms for the review. Then, we selected appropriate databases and screened them based on inclusion and exclusion criteria. And we completed the title-abstract and full-read step to obtain the final selected papers. At the same time, we also adopted the "snowballing" approach in the step of full reading to extract more relevant information~\cite{wohlin2014guidelines}.

\textbf{Search String.} We identified the main search terms combined with the research questions. Therefore, we started with the terms \emph{``sixth generation''} and \emph{``software engineering''} because the field of our research is about 6G Software Engineering. Secondly, we found common alternative spellings and synonyms for both terms \emph{``sixth-generation''},  \emph{``6g''},  \emph{``software''} and  \emph{``software development''} to ensure the consistency and completeness. Thirdly, we selected \emph{``communicat*''}, \emph{``network*''}, \emph{``techn*''} and \emph{``cellular''} to delineate main subject areas. Finally, we used boolean operators to relate the various terms identified so far:

\begin{center}
    \emph{``(sixth-generation OR 6g OR ``sixth generation'')''}
    \\\emph{AND}
    \\\emph{``(software OR ``software development'' OR ``software engineering'')''}
    \\\emph{AND}
    \\\emph{``(communicat* OR network* OR techn* OR cellular)''}
\end{center}

\textbf{Bibliographic Sources.} After determining the search criteria, we defined the bibliographic sources to search for. We selected Scopus, IEEE Xplore, Association for Computing Machinery(ACM) digital library and Web of Science(WoS). The selection of these databases mainly depended on their popularity and completeness, and they are widely recognized as the most representative of research in software engineering~\cite{kitchenham2007guidelines}. In addition, we decided not to rely on Google Scholar in order to avoid including papers that are not peer-reviewed (e.g., ArXiv and others).

\textbf{Inclusion and Exclusion Criteria.} For inclusion criteria, it was related to the usefulness of a paper. We considered the defined research questions and RQ$_{1.1}$ was the core and the rest of the research questions were based on it. As described in Table~\ref{tab:incexc}, we included papers based on the criteria that map to RQ$_{1.1}$.

For the exclusion criteria, we first filtered out papers that were not written in English, were repetitive, and were not related to our research questions. In addition, we excluded papers that were not peer-reviewed or papers proposing new methods that were not validated. Finally, we excluded papers that were not accessible by the institution.

\begin{table}[!ht]
\centering
\caption{Inclusion and Exclusion Criteria} 
\label{tab:incexc}
\resizebox{\linewidth}{!}{
\begin{tabular}{l|p{7cm}}
\toprule
\textbf{Inc./Exc.} & \textbf{Criteria}\\ 
\midrule
Inclusion & Papers mention on 6G Software Development, Process, Orchestration and Offloading Methods, Software Architecture\\
\hline
Exclusion & Not in English  \\
& Duplicated/extension has been included \\
& Out of topic \\
& Non peer-reviewed papers \\
& Not accessible by institution \\ 
& Position papers that present new method without validation \\
\bottomrule
\end{tabular}}
\end{table}

\textbf{Search Process.} After defining the search terms, databases and inclusion/exclusion criteria, we started the application of the search strings on the search databases. The initial literature search numerical results are displayed in Table~\ref{tab:library}. The initial candidate set was composed of 865 papers, which was reduced to 479 papers after removing the duplicates.

\begin{table}
\centering
% \scriptsize
\caption{Initial Literature Search by Library} 
\label{tab:library} 
\adjustbox{max width=\linewidth}{
\begin{tabular}{l|l}
\toprule
\textbf{Library} & \textbf{Count}
\\\midrule
Scopus & 446 \\
IEEE & 192 \\ 
ACM & 11 \\ 
Web of Science & 216 \\ \hline
Non-duplicates & 479 \\
\bottomrule
\end{tabular}
}
\end{table}

Then we applied the inclusion and exclusion criteria to each paper. Each paper was read and evaluated by %the first 
two authors, and in the case of disagreement, the third author decided whether to include or exclude these papers. Of the initial 479 papers, 88 were included by reading title and abstract. The second step in our selection process was a full reading of the papers, with %the first
two authors deciding in each case whether or not to be included, and in this process, in 10 cases a third author was needed to resolve the disagreement. The next step is snowballing. The references of the currently included papers are also analyzed according to the inclusion criteria. Finally, 18 papers were included.

\subsection{Data Extraction}

To summarize the included papers, we extracted relevant data information when doing the full read part. In order to better distinguish between the different knowledge areas in 6G Software Engineering to answer the three research questions, we divided the related fields into three main areas, based on the  Work Packages(WP) of the 6G Soft Project:

\begin{itemize}

    \item \textbf{WP1:} Software development methods, and processes.\\ 
    In this area, we considered the SWEBOK software engineering areas~\cite{swebok}: 
Software Requirements, Software Construction, Software Testing, Software Maintenance, Software Configuration Management, Software Engineering Management, Software Engineering Process, Software Engineering Models and Methods, Software Quality. 

    \item \textbf{WP2:} Software architecture

    \item \textbf{WP3:} Orchestration and offloading methods

    % \item \textbf{WP4:} Business-driven software development
    
\end{itemize}

To screen each paper, we assigned two researchers to independently review them. To ensure fairness, we mixed up the assignments and made sure each researcher had a similar number of papers to review with other members of the team. In case of disagreement, a third author was brought in to reach a consensus. This was done to improve the reliability of our study. Finally, to evaluate the inter-rater agreement before involving the third author, we calculated Cohen's kappa coefficient. The Cohen's $K$ coefficient scored 0.71, which may be interpreted as good.

% We also set up the label of ``topic type''. When reading the full paper, the authors extracted useful data and categorized it according to WP and ``topic type'', and this helped in answering RQ$_{1.1}$ and RQ$_{1.2}$. In addition, we set up the label of ``result type'' that was used to extract the results of papers, and this helped in answering RQ$_{2.1}$ and RQ$_{2.2}$.

\section{Results}

The following section presents the results of our mapping study. 
% Each of the following paragraphs is aimed at answering the main topics proposed by our research questions. 
The selected papers(SPS) can be found in APPENDIX.

\subsection{RQ$_1$ Main topics and studies in 6G Software Engineering}

\begin{figure}[h]
    \centering
    \includegraphics[width=\linewidth]{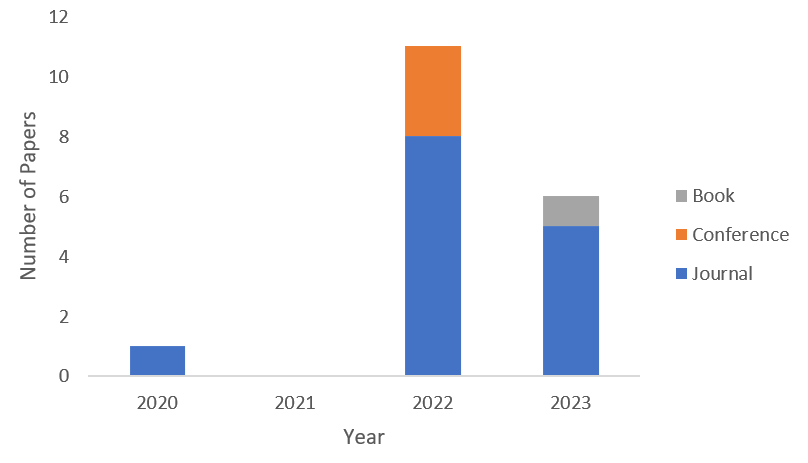}
    \caption{Number of publications over the years}
    \label{fig:cumulativeover}
\end{figure}

\textbf{RQ1.1} According to the literature review, 18 papers are finally included that have been proposed in 6G Software Engineering. Fig.~\ref{fig:cumulativeover} depicts the number of papers published over the years: 1 journal published in 2020, 11 papers with 8 journals and 3 conferences in 2022, and 6 papers with 5 journals and 1 book chapter in 2023. The figure shows that the number of publications has increased, with the first publication dating back to 2020 because 5G became commercially available in 2020, and most researchers started research on 6G after this. At the same time, Fig.2 also clearly presents the document type of the published papers, showing that the vast majority of publications are in journals, with conference papers and books in the minority.
% Then we have 11 papers in 2022 and 6 in 2023. The cumulative number of papers published in this field has also been increasing, with one paper published at the end of 2020, 12 papers at the end of 2022, and reaching a total of 18 papers by July 2023.

% \begin{figure}[h]
%     \centering
%     \includegraphics[width=\linewidth]{FIG2.png}
%     \caption{Publication over years}
%     \label{fig:publicationoveryears}
% \end{figure}

It is worth noting that there is no paper published in 2021 related to 6G Software Engineering. It might be due to the fact that researchers gradually started researching 6G for a short period of time and are still in the early stages. Hence, there are almost no papers related to 6G Software Engineering output in 2021 and 2022. However, since 2022, there has been a speedy increase in published papers. This indicates that the research of the previous two years has begun to bring results in the form of papers. It is also important to note that we considered the paper published up to July 2023. Therefore, data about the year 2023 might not include all the papers published in that year.

\begin{figure}
    \centering
    \includegraphics[width=\linewidth]{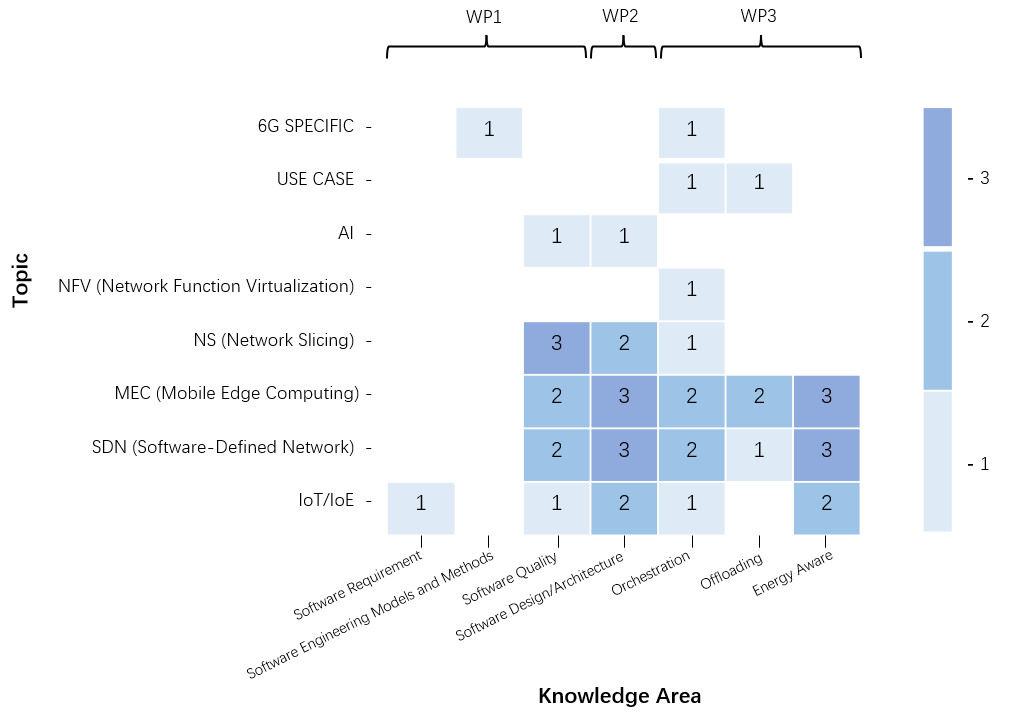}
    \caption{Distribution of selected papers by topic and knowledge area}
    \label{fig:distribution}
\end{figure}

\textbf{RQ1.2} Fig.~\ref{fig:distribution} gives insights into the variation in the distribution of knowledge areas across the different topics related to 6G Software Engineering. In the figure, the number of papers is represented by color darkness, with darker colors indicating a higher number. It is possible to see that 6G Software Engineering papers focus on WP1(Software Requirement~\ref{SP3}, Software Engineering Models and Methods~\ref{SP7}, and Software Quality~\ref{SP1}~\ref{SP4}~\ref{SP8}~\ref{SP13}~\ref{SP14}~\ref{SP15}, WP2~\ref{SP1}~\ref{SP2}~\ref{SP3}~\ref{SP8}~\ref{SP9}~\ref{SP17}~\ref{SP18}) and WP3 (Orchestration~\ref{SP1} ~\ref{SP5} ~\ref{SP6} ~\ref{SP7} ~\ref{SP9} ~\ref{SP10}~\ref{SP16}, Offloading~\ref{SP1}~\ref{SP2}~\ref{SP11} and Energy Aware~\ref{SP2}~\ref{SP4}~\ref{SP6}~\ref{SP12}). Therefore, we found that software architecture is the area that has received the most attention in 6G Software Engineering. Additionally, these included papers cover topics related to 6G Software Engineering are IoT/IoE, SDN(Software-Defined Network), MEC(Mobile Edge Computing), NS(Network Slicing), NFV(Network Function Virtualization), AI, Use Case and 6G Specific.

SDN is the topic that has received the most attention in 6G Software Engineering, with 6 papers~\ref{SP2}~\ref{SP4}~\ref{SP5}~\ref{SP8}~\ref{SP12}~\ref{SP18}. Then IoT/IoE~\ref{SP3}~\ref{SP4}~\ref{SP5}~\ref{SP12}~\ref{SP17} and MEC~\ref{SP1}~\ref{SP2}~\ref{SP4}~\ref{SP6} \ref{SP17}, with 5 papers each. In addition, NS~\ref{SP8}~\ref{SP9}~\ref{SP13}~\ref{SP15}, AI~\ref{SP14}~\ref{SP17}, and Use case~\ref{SP10}~\ref{SP11} have received less attention. Finally, NFV~\ref{SP16} and 6G Specific~\ref{SP7} have received the least attention, with only one paper selected for each.

Based on Fig.~\ref{fig:distribution}, we found that IoT/IoE, SDN and MEC are the main research topics in 6G Software Engineering, and these focus topics have a relatively even distribution of papers across different knowledge areas, while other topics have a little and scattered distribution(It's worth noting one paper exists that contains multiple topics and WPS). For example, the most focused topic of SDN has a relatively even distribution of papers across the different knowledge areas, with two or three papers for each of the knowledge areas considered (except for Software Requirement and Software Engineering Models and Methods). In contrast, the topic of NFV only has one paper on the knowledge area of Orchestration.

\subsection{RQ$_2$ Main results obtained in 6G Software Engineering}

\begin{figure}[h]
    \centering
    \includegraphics[width=\linewidth]{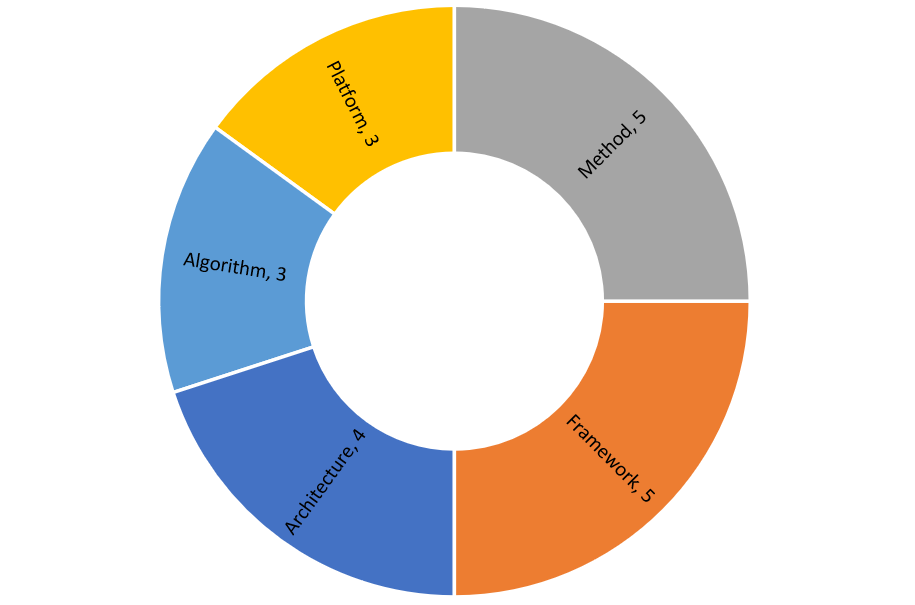}
    \caption{Distribution of the main result types}
    \label{fig:resulttypes}
\end{figure}

\textbf{RQ2.1} Fig.~\ref{fig:resulttypes} shows the result types achieved by the included papers. The most frequent result is ``Method'' and ``Framework'', with 5 papers reporting it for each. Al-Hammadi and Islam~\ref{SP2} implement a collaborative offloading method among MEC servers based on the edge server’s resources and neighbors’ status to alleviate network congestion, and formulate a hierarchy SDN-powered MEC network framework comprising three tiers in the method. Habibi et al.~\ref{SP7} provide guidelines on how the novel software building blocks can be integrated and deployed as part of a DevOps workflow, and propose a Service Management and Orchestration (M\&O) framework for 6G at the same time. Alotaibi and Barnawi~\ref{SP11} propose IDSoft, a novel softwarized solution that resides across the network infrastructure and leverages 6G enabling technologies. Shukla et al.~\ref{SP12} propose 6G-SDI, a Software-Defined Network (SDN)-based green communication method for 6G-enabled Internet of Things (IoT) to control real-time actuation and flow-table configuration. Abdulqadder and Zhou~\ref{SP13} tackle issues such as security, QoS, and resource consumption issues through related network slicing and load balancing methods in SDN/NFV assisted 6G environments. In addition, except for the first two papers that propose the framework in the methods, there are three other papers that present the framework separately. Meenakshi et al.~\ref{SP4} use a framework constructing a design of systematic Wireless Inventory trackers (WIT) using heterogeneous IoT(HIoT) networks over 6G Computing and MEC/SDWAN for improving the prolonged lifetime and low energy consumption for efficient communication in 6G network. Manogaran et al.~\ref{SP5} propose the service virtualization and flow management framework (SVFMF) for the reliable utilization of resources in the 6G-cloud environment. Janbi et al.~\ref{SP17} propose a framework for Distributed AI as a Service (DAIaaS) provisioning for Internet of Everything (IoE) and 6G environments.

In addition, the result type ``Architecture'' and ``Platform'' has 4 papers reporting it as the main achievement. Bojović et al.~\ref{SP8} design a multi-slice architecture to develop highly flexible dynamic queue management software and moved it completely to the application layer. Alonso-Lupez et al.~\ref{SP9} propose an architecture that will enable privacy-aware slicing and security service orchestration. Moreover, Alotaibi and Barnawi~\ref{SP11} propose HFL (hierarchical FL) architecture that has an additional offloading mechanism to enhance and evaluate performance in terms of relevant FL aspects, such as accuracy, communication efficiency, and convergence. Tao et al.~\ref{SP18} propose a novel software-defined DTN architecture with digital twin function virtualization (DTFV) for adaptive 6G service response. On the other hand, for the results of ``Platform'', Ateya et al.~\ref{SP1} develop a MEC platform and introduce a seamless migration for complex 5G/6G tasks. Furthermore, Katiyar et al.~\ref{SP3} develop an IoT platform and focus on its middleware layer connection with other layers. Cao et al.~\ref{SP10} propose C-ITS, a Cooperative intelligent transport system that allows softened resource management and allocation in 6G networks with autonomy and smart sensing. Different from the previous, Kamruzzaman and Alruwaili~\ref{SP14} propose a system and use a new technology named Optimizing Computer Vision with AI-enabled technology (OCV-AI model) to address the issues mentioned and improve the system's outcome.

Finally, the result for ``Algorithm'' is given in three papers. Ajibola et al.~\ref{SP6} propose a heuristic for energy-efficient and delay-aware placement (HEEDAP) for applications in fog networks (an algorithm). Cao et al.~\ref{SP15} propose an efficient resource allocation algorithm, labeled TailoredSlice-6G, so as to implement the tailored resource allocation of slices in 6G networks and they did a comprehensive simulation to emphasize the benefits of TailoredSlice-6G, results showing that it outperforms representative heuristics in the literature. Ye et.al~\ref{SP16} propose a heuristic decoupled SFC orchestration algorithm (HDSFCO) with low complexity to minimize the overall resource costs, fully considering the time evolution characteristics of dynamic network topology.

\textbf{RQ2.2} According to the results in RQ$_{2.1}$, we found that there are only 5 papers focusing on the result of ``framework'' ~\ref{SP2}~\ref{SP4}~\ref{SP5} \ref{SP7}~\ref{SP17}. However, these frameworks are newly proposed and not duplicated. This represents a total of five tools/frameworks that have been proposed in the selected 18 papers in the field of 6G Software Engineering, but they are not currently studied by other researchers.

\subsection{RQ$_3$ Main authors and distribution in 6G Software Engineering}

\begin{figure}[t]
    \centering
    \includegraphics[width=\linewidth]{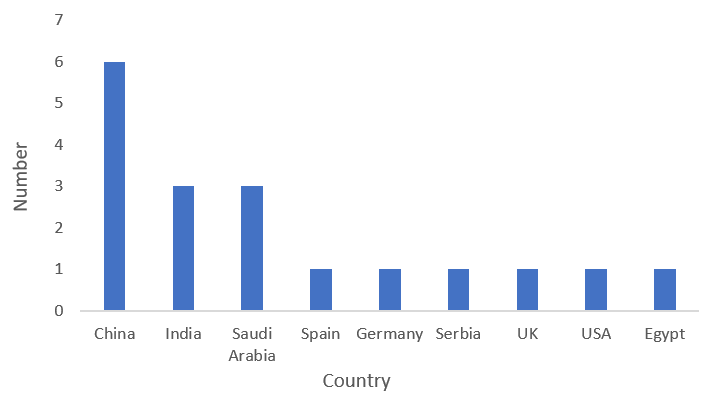}
    \caption{Number of countries of the first authors}
    \label{fig:year}
\end{figure}

\begin{figure}[t]
    \centering
    \includegraphics[width=\linewidth]{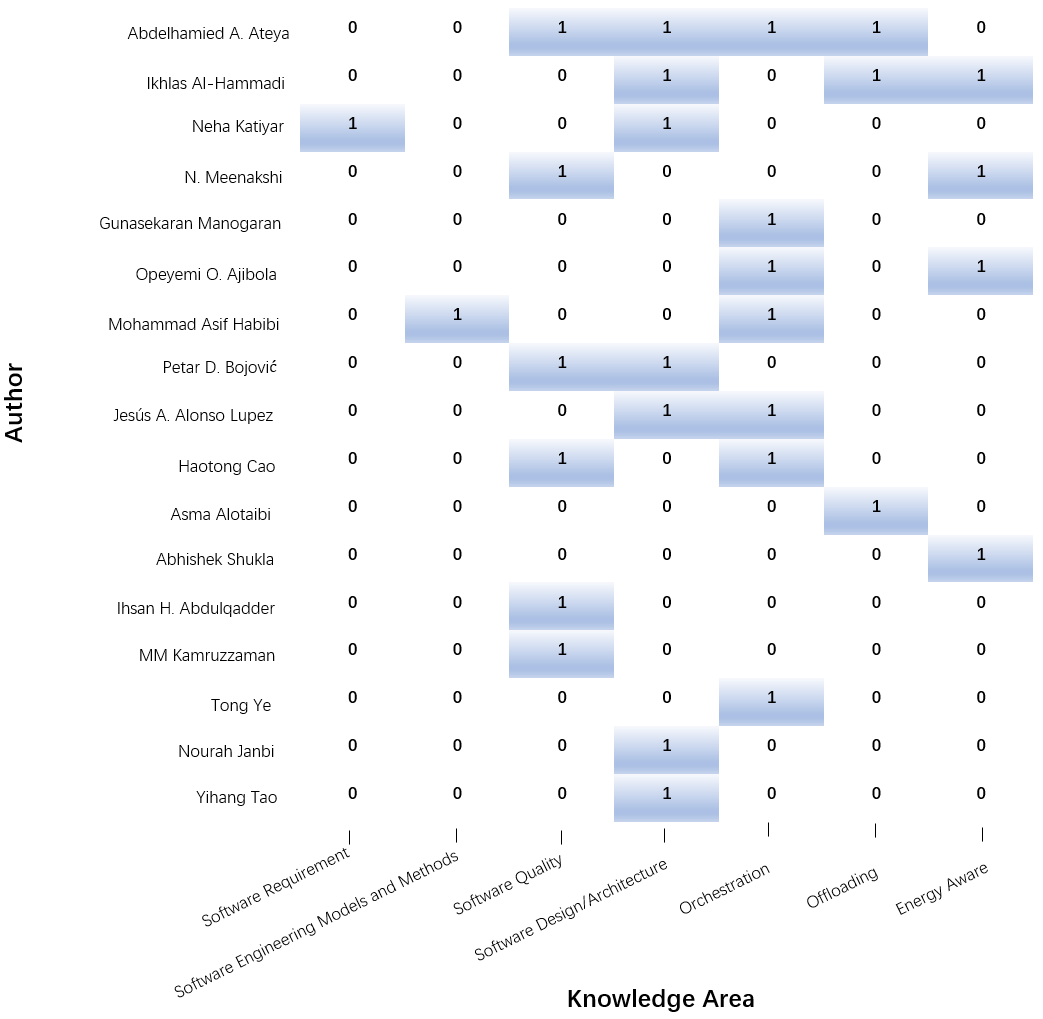}
    \caption{Distribution of first authors for each considered knowledge area}
    \label{fig:distributionyears}
\end{figure}

\textbf{RQ3.1} We counted the authors in the selected 18 papers, there are 86 authors in total, including 17 first authors. Fig.~\ref{fig:year} depicts the number of countries for the first authors, showing that the first authors are located in nine countries: China, India, Saudi Arabia, Spain, Germany, Serbia, the UK, the USA and Egypt. Among them, China has the highest number of first authors with six, followed by India and Saudi Arabia with three each. The other six countries all have only one first author. In addition, the main researchers (first authors' names) are shown in Fig.\ref{fig:distributionyears}.

\textbf{RQ3.2} Fig.~\ref{fig:distributionyears} presents the distribution of first authors for each considered knowledge area. We found that among these first authors, Abdelhamied A. Ateya's paper contains the most knowledge areas, including Software Quality, Software Design/Architecture, Orchestration, and Offloading. Followed by IKlas, whose paper contains three knowledge areas. The other authors' papers contain only one or two knowledge areas. Interestingly, almost all the authors have publications focused on a restricted number of knowledge areas.

\section{Discussion}

The analysis of the literature shows that although researchers and practitioners frequently mention the field of 6G Software Engineering, there is only a limited number of research papers investigating software process, software architecture, orchestration and offloading methods.

\begin{figure*}
    \centering
    \includegraphics[width=0.7\textwidth, height=7.5cm]{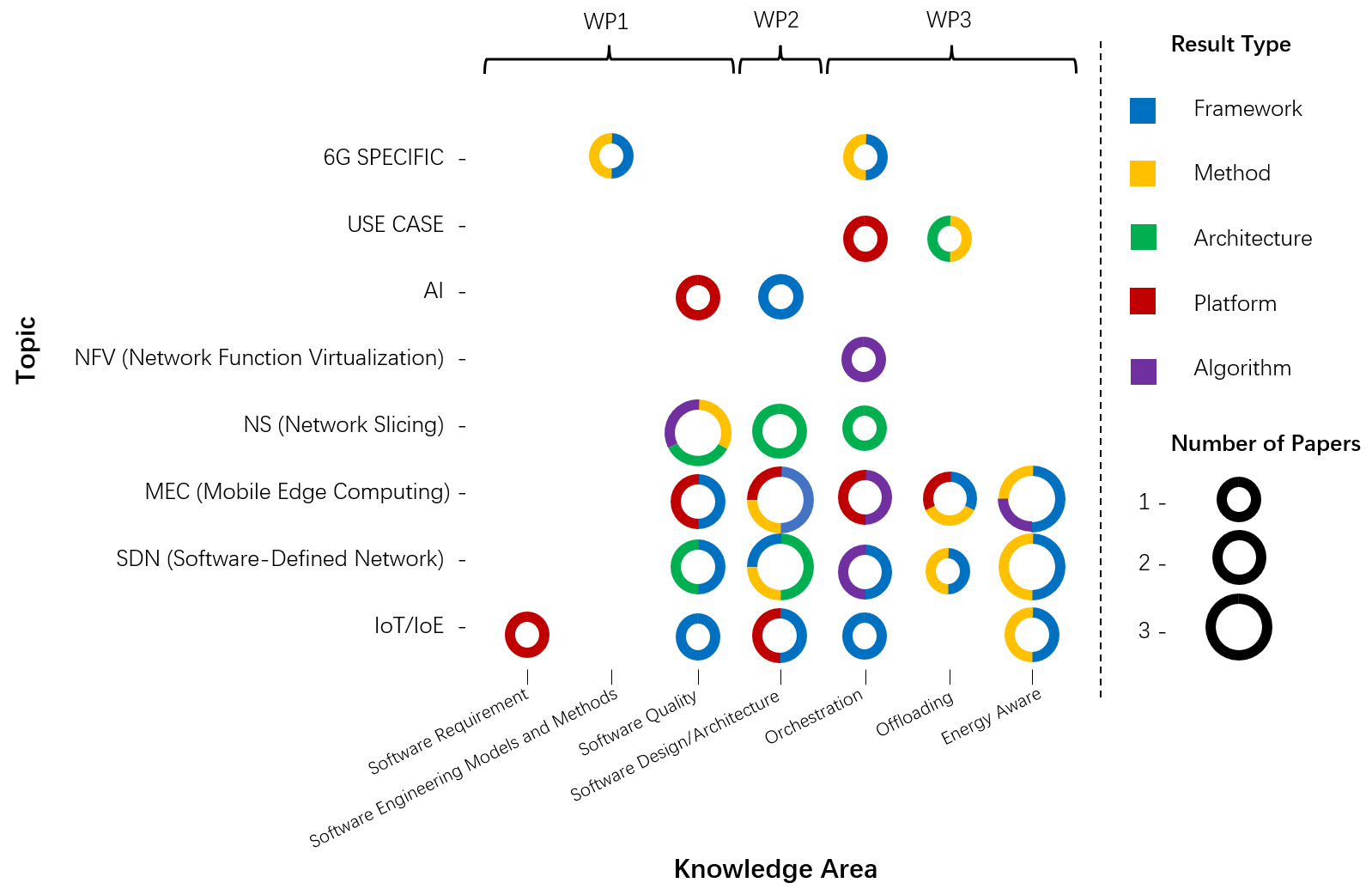}
    \caption{Distribution of selected papers by topic, knowledge area and result type}
    \label{fig:Comprehensive}
\end{figure*}

As followed in Fig.~\ref{fig:Comprehensive}, we considered combining RQ$_1$ and RQ$_2$ in our analysis and obtained the differences in the distribution of knowledge areas and result types for the different topics related to 6G Software Engineering. In the figure, the number of papers is represented by the size of the circles and the result types are respectively presented in different colors. We found from the figure that the four areas of ``Software Quality-NS'', ``Software Architecture-SDN'', ``Software Architecture-MEC'' and ``Energy Aware-MEC'' have the richest result types and the highest number of papers. This suggests that researchers have focused on these four cross-cutting areas of research. However, there is almost a gap in research on software requirements and software engineering models and methods. Moreover, it is worth noting that in Energy Aware, frameworks and methods are the predominant types of paper results and only one paper involves algorithms.

The limitation of this study is that currently the various areas related to 6G are still in the research stage and have not yet been standardized and commercialized. Many of the researchers are rapidly progressing their research but have not published it, so the statistics on the results are only staged in the early stage and not final. In addition, since 6G involves a number of fields, some related studies in 6G are actually at the intersection of software engineering and other fields, so researchers need to consider not only the software engineering field but also not ignore the fields that intersect with it.

Regarding the implication, this study provides insight into the existing research topics closely related to 6G software and their relation to software engineering knowledge areas. It also discovers the potential research gaps in the missing software engineering theoretical support in critical 6G subdomains. Furthermore, researchers and practitioners shall also know what results (e.g., frameworks, methods, platforms, etc.) exist and what is still required. These will help researchers be informed about the current state of research in this field so that they have a basis on which to continue their study. At the same time, these may also be helpful for novices and researchers who are interested in joining this field of research to discover related areas and potential collaborators. Finally, these may be useful for practitioners to understand how 6G has been integrated with Software Engineering.
For future work, we shall focus on further investigating the critical aspects of 6G software and exploring the domain knowledge of software engineering for intersections like edge computing.

\section{Threats to Validity}

By following the guidelines proposed by Ampatzoglou et al. \cite{ampatzoglou2019identifying}, we discuss the potential threats to the validity of this study. This guideline is proposed especially for secondary studies in the software engineering domain. Herein, we discuss the Study Selection Validity, Data Validity, and Research Validity of our study and the potential mitigation of their impact.

\textbf{Study Selection Validity.} Regarding the search strategy, review protocol, and data extraction steps of this secondary study, we followed strictly the guidelines proposed by Petersen et al. \cite{petersen2008systematic,petersen2015guidelines}. Guided by the well-recognized studies, we largely mitigated the threats to the initial search and filtering processes within the planning phase. Considering the criticality of the search string for this study, we carefully formulate it by including all the keywords identified from research questions and diversifying using synonyms. Though searching thusly covers most of our target publications; however, it is still inevitable that potential issues may arise during the search process when critical information does not appear in the publication's titles or abstracts. To mitigate such limitations and to extend the coverage of studies, we used snowballing as complementary by checking all the references listed in the selected studies and also all the papers citing the selected ones. Both the inclusion and exclusion criteria sets are also following the guidelines proposed by Petersen et al. \cite{petersen2008systematic}. The criteria comply closely with the research questions and the target research theme and are also validated by all authors. Furthermore, for both filtering steps of "title-abstract" and "full-paper" reading, two authors conducted the reading and filtering independently which largely reduced the selection bias. Regarding the disagreements in the process, the third author, who is also a domain expert, was invited to break the tie. Meanwhile, the study selection was conducted in August 2023 which is, to a large extent, timely and reduces the threat of incompleteness. 

\textbf{Data Validity.} Within the process of data extraction, two authors conducted the extraction work independently. The process is iterative and based on the open coding method to identify the classification schema. For specific research questions, the established categories can be found in well-recognized literature. For example, for RQ1, we adopt the categories for main topics in the software engineering domain from the Software Engineering Body of Knowledge (SWEBOK) \cite{bourque2004swebok}. The guide is an emerging consensus of software engineering domain knowledge with also explicitly defined categories of topics that are also commonly adopted by other studies. For RQ2, the set of categories is extracted using the open coding method which largely reduced the bias in classification schema and the mapping of data. With the categories clearly defined, the corresponding results can be easily summarized and presented in different formats of charts. However, the potential bias in the publication is an inevitable threat. Due to confidential policies, the selection of literature is from open-accessible academic papers when the experiences and knowledge from the industry are limited. Such a threat to validity can be mitigated in future studies via industrial surveys.

\textbf{Research Validity.} The research method is determined via multiple rounds of discussion among all authors before the start of this study. The decision on adopting a systematic mapping study was agreed upon by all authors. Such a decision-making process shall mitigate the threat of research method bias. After the selection of the research method, all the authors also determined the research question together via several iterations. On the other hand, the study can be replicated by following the replication documentation and the steps meticulously. The search string and the other details on the process of the systematic mapping study are also explicitly displayed therein. Accompanied with the advance of technology in the future, it is highly likely that such replication is applied to investigate the network and software architecture and design for the next generations.

\section{Conclusion}

In this systematic mapping study, we aimed to provide a comprehensive overview of the current research status of 6G Software Engineering. We conducted a systematic mapping study to investigate the current research status of 6G Software Engineering, especially in software process, software architecture, orchestration and offloading methods. We surveyed 479 research papers and identified 18 relevant studies. Through an extensive analysis of these 18 studies, the results show that current studies focus on 6G software quality, software architecture, and orchestration \& offloading methods. Of these, software architecture and SDN(Software-Defined Networks) are respectively areas and topics that have received the most attention in 6G software engineering. In addition, many existing studies focus on NS, MEC, and SDN of 6G networks when the other topics, e.g., AI and NFV, need to be more studies. Finally, the main authors of these research papers are mainly located in China, India and Saudi Arabia.

This study provides insights into the critical aspects of 6G software engineering in terms of the different knowledge areas and can inform researchers about the existing research status in this domain so that they can have a basis when continuing research in this field. In addition, we also provide a systematic mapping of the distribution of researchers who are currently focusing on 6G software engineering, which may be useful to newcomers and researchers interested in getting started in this research field to discover relevant areas and potential collaborators.
Finally, it provides a comprehensive synthesis and analysis of the research conducted in the field of 6G Software Engineering, which might be useful to researchers and practitioners to learn how 6G and Software Engineering have been combined so far. 

Our future research agenda is based on the main findings of our systematic mapping study. We shall focus on investigating the critical aspects of 6G software and exploring the domain knowledge of software engineering for intersections like edge computing. We plan to survey some people (companies) engaged in 6G software engineering and conduct further analysis of the architectural framework in 6G mentioned in RQ2.

%\begin{acks}
  %This work was supported by a grant from the Academy of Finland (grant n. 349488 - MuFAno) and a grant from Business Finland ("6G-Bridge 6GSoft").

  %The authors would like to thank Dr. Yuhua Li for providing the matlab code of  the \textit{BEPS} method. 

  %The authors would also like to thank the anonymous referees for their valuable comments and helpful suggestions. The work is supported by the \grantsponsor{GS501100001809}{National Natural Science Foundation of China}{http://dx.doi.org/10.13039/501100001809} under Grant No.:~\grantnum{GS501100001809}{61273304} and~\grantnum[http://www.nnsf.cn/youngscientsts]{GS501100001809}{Young Scientsts' Support Program}.

%\end{acks}

\bibliographystyle{ACM-Reference-Format}
\bibliography{bibliography}

% \newpage

\appendix
%Appendix A
\section*{Appendix A: The Selected studies} 
\label{The Selected Papers}

{\small
 \begin{enumerate} [labelindent=-5pt,label={[S}{\arabic*]}]

\item \label{SP1}
Ateya, A. A., Alhussan, A. A., Abdallah, H. A., Khakimov, A., \& Muthanna, A. (2023). Edge Computing Platform with Efficient Migration Scheme for 5G/6G Networks. Computer Systems Science \& Engineering, 45(2).

\item \label{SP2}
Al-Hammadi, I., Li, M., \& Islam, S. (2023). Independent tasks scheduling of collaborative computation offloading for SDN-powered MEC on 6G networks. Soft Computing, 27(14), 9593-9617.

\item \label{SP3}
Katiyar, N., Kumari, P., Sakhshi, S., \& Srivastava, J. (2023). 9 Trending IoT Platforms. Intelligent Analytics for Industry 4.0 Applications, 131.

\item \label{SP4}
Meenakshi, N., Jaber, M. M., Pradhan, R., Kamruzzaman, M. M., Maragatham, T., Ramamoorthi, J. S., \& Murugesan, M. (2023). Design systematic wireless inventory trackers with prolonged lifetime and low energy consumption in future 6G network. Wireless Networks, 1-22.

\item \label{SP5}
Manogaran, G., Baabdullah, T., Rawat, D. B., \& Shakeel, P. M. (2021). AI-assisted service virtualization and flow management framework for 6G-enabled cloud-software-defined network-based IoT. IEEE Internet of Things Journal, 9(16), 14644-14654.

\item \label{SP6}
Ajibola, O. O., El-Gorashi, T. E., \& Elmirghani, J. M. (2022). Disaggregation for energy efficient fog in future 6G networks. IEEE Transactions on Green Communications and Networking, 6(3), 1697-1722.

\item \label{SP7}
Habibi, M. A., Sánchez, A. G., Pavón, I. L., Han, B., Landi, G., Sayadi, B., ... \& Virdis, A. (2022, October). Enabling Network and Service Programmability in 6G Mobile Communication Systems. In 2022 IEEE Future Networks World Forum (FNWF) (pp. 320-327). IEEE.

\item \label{SP8}
Bojović, P. D., Malbašić, T., Vujošević, D., Martić, G., \& Bojović, Ž. (2022). Dynamic QoS management for a flexible 5G/6G network core: a step toward a higher programmability. Sensors, 22(8), 2849.

\item \label{SP9}
Alonso-Lupez, J. A., Hernández, L. A. M., Arteaga, S. P., Orozco, A. L. S., Villalba, L. J. G., Pastor, A., \& García, D. L. (2022, July). Level of Trust and Privacy Management in 6G Intent-based Networks for Vertical Scenarios. In 2022 1st International Conference on 6G Networking (6GNet) (pp. 1-4). IEEE.

\item \label{SP10}
Cao, H., Garg, S., Kaddoum, G., Singh, S., \& Hossain, M. S. (2022). Softwarized resource management and allocation with autonomous awareness for 6G-enabled cooperative intelligent transportation systems. IEEE Transactions on Intelligent Transportation Systems, 23(12), 24662-24671.

\item \label{SP11}
Alotaibi, A., \& Barnawi, A. (2023). IDSoft: A federated and softwarized intrusion detection framework for massive internet of things in 6G network. Journal of King Saud University-Computer and Information Sciences, 35(6), 101575.

\item \label{SP12}
Shukla, A., Ahmed, N., Roy, A., \& Misra, S. C. (2022). Softwarized management of 6G network for green Internet of Things. Computer Communications, 187, 103-114.

\item \label{SP13}
Abdulqadder, I. H., \& Zhou, S. (2022). SliceBlock: context-aware authentication handover and secure network slicing using DAG-blockchain in edge-assisted SDN/NFV-6G environment. IEEE Internet of Things Journal, 9(18), 18079-18097.

\item \label{SP14}
Kamruzzaman, M. M., \& Alruwaili, O. (2022). AI-based computer vision using deep learning in 6G wireless networks. Computers and Electrical Engineering, 102, 108233.

\item \label{SP15}
Cao, H., Du, J., Zhao, H., Luo, D. X., Kumar, N., Yang, L., \& Yu, F. R. (2021). Toward tailored resource allocation of slices in 6G networks with softwarization and virtualization. IEEE Internet of Things Journal, 9(9), 6623-6637.

\item \label{SP16}
Ye, T., Zhang, J., Zhao, C., Tang, Y., \& Zhu, C. (2022, October). Service Function Chain Orchestration in 6G Software Defined Satellite-Ground Integrated Networks. In 2022 6th International Conference on Communication and Information Systems (ICCIS) (pp. 71-76). IEEE.

\item \label{SP17}
Janbi, N., Katib, I., Albeshri, A., \& Mehmood, R. (2020). Distributed artificial intelligence-as-a-service (DAIaaS) for smarter IoE and 6G environments. Sensors, 20(20), 5796.

\item \label{SP18}
Tao, Y., Wu, J., Lin, X., \& Yang, W. (2023). DRL-Driven Digital Twin Function Virtualization for Adaptive Service Response in 6G Networks. IEEE Networking Letters.

\end{enumerate}}

\end{document}